\documentclass[aps,showpacs,pre,onecolumn]{revtex4}
\usepackage{amsmath}
\usepackage{graphicx,natbib,amssymb}
\usepackage{setspace}
\usepackage{color}
\begin{document}

\title{What kind of noise is brain noise: Anomalous scaling behavior of the resting brain activity fluctuations.\footnote{Cite as: Fraiman D. and Chialvo D.R. (2012) What kind of noise is brain noise: anomalous scaling behavior of the resting brain activity fluctuations. Front. Physio. 3:307. doi: 10.3389/fphys.2012.00307}}

\author{ Daniel Fraiman$^{1,2}$ \& Dante R. Chialvo$^{1,3,4}$}

\affiliation{$^1$ Consejo Nacional de Investigaciones Cient\'ificas y Tecnol\'ogicas (CONICET), Buenos Aires, Argentina.}
\affiliation {$^2$ Departamento de Matem\'atica y Ciencias,
Universidad de San Andr\'es, Argentina.}
\affiliation{$^3$Facultad de Ciencias M\'edicas, Universidad Nacional de Rosario, Rosario, Argentina}
\affiliation{$^4$ David Geffen School of Medicine, University of California, Los Angeles, CA, USA.}

\begin{abstract} 
The study of spontaneous fluctuations of brain activity,  often referred as brain noise, is getting increasing attention in functional magnetic resonance imaging (fMRI) studies. Despite important efforts, much of the statistical properties of such fluctuations remain largely unknown. This work scrutinize these fluctuations looking at specific statistical properties which are  relevant to clarify its dynamical origins.
Here, three statistical features which clearly differentiate brain data from naive expectations for random processes are uncovered:  
First, the variance of the fMRI mean signal as a function of the number of averaged voxels  remains constant across a wide range of observed clusters sizes.  
Second,  the anomalous behavior of the variance is originated by bursts of synchronized activity across regions, regardless of their widely different sizes. 
Finally, the correlation length (i.e., the length at which the correlation strength between two regions vanishes) as well as mutual information diverges with the cluster's size considered, such that arbitrarily large clusters exhibit the same collective dynamics than smaller ones.  
These three properties are known to be exclusive of complex systems exhibiting critical dynamics, where the spatio-temporal dynamics show these peculiar type of fluctuations. Thus,  these findings are fully consistent with previous reports of brain critical dynamics, and are relevant for the interpretation of the role of fluctuations and variability in brain function in health and disease.  

\end{abstract}


\maketitle

\section{Introduction}
It is now recognized that important information can be extracted from the brain spontaneous activity, as exposed by recent analysis \cite{biswal,fox2007,Beckmann-2009}. For instance, the structure and location of large-scale brain networks can be derived from the interaction of cortical regions during rest which closely match the same regions responding to a wide variety of different activation conditions \cite{fox2007,Beckmann-2009}. These so-called resting state networks (RSN) can be reliably computed from the fluctuations of the blood oxygenated level dependent signal (BOLD)  signals of the resting brain, with great consistency across subjects \cite{xiong,cordes,beckmann2005} even during sleep \cite{fuku} or anesthesia \cite{vincent}.

In the same direction, the information content of the brain BOLD signal's variability {\it per se} is receiving increasing interest. Recently \cite{noise} it was shown in  a group of subjects of different age, that the BOLD signal variability (standard deviation) is a better predictor of the subject age than the average.  Furthermore, additional work focused on the relation between the fMRI signal variability and a task performance, concluded that faster and more consistent performers exhibit significantly higher brain variability across tasks than the poorer performing subjects \cite{Garrett}. Overall, these  results suggest that the understanding of the brain resting dynamics can benefit from a detailed study of the BOLD variability {\it per se}.  

In this work we  characterize the statistical properties of the spontaneous BOLD fluctuations and discuss its possible dynamical mechanisms. The paper is organized as follow: In the next section the origin of the data is described as well the preprocessing of  the signal. The definitions of regions of interest is described as well as how to construct subsets of different sizes, needed to compute fluctuations. The results section starts with  the analysis of the average spontaneous fluctuations for each resting state network, which identify anomalous scaling of the variance as a function of the number of elements. Next, this anomaly is explored to determine its origins by studying in detail the temporal correlations in clusters of different sizes. Finally the analysis of the correlation length is described, revealing a distinctive divergence with the size of the cluster considered. The paper close with  a  discussion of the relevance of the uncovered anomalous scaling for the current views of large scale brain dynamics. For clarity of presentation, the calculations that are not central to the main message of the paper, are presented separately in an Appendix.
%%% 
\section{Methods}

\emph{\bf Data Acquisition.} fMRI data was obtained from five healthy right-handed subjects (21-60 years old, mean=40.2) using a 3T Siemens Trio whole-body scanner with echo-planar imaging capability and the standard radio-frequency head coil.  Subjects were scanned following a typical brain resting state protocol \cite{fox2007} lying in the scanner and asked to keep their mind blank, eyes closed and avoid falling asleep. All participants gave written informed consent to procedures approved by the IRB Committee of the University of Islas Baleares (Mallorca, Spain) who approved the study.

\emph{\bf Image pre-processing and analysis.}
In each subject, 240 BOLD images, spaced by 2.5 sec., were obtained from 64x64x49  voxels of dimension 3.4375mm x 3.4375mm x 3mm.
Preprocessing was performed using FMRIB Expert Analysis Tool (FEAT, \cite{jezzard}, http://www.fmrib.ox.ac.uk/fsl), involving motion correction using MCFLIRT; slice-timing correction using Fourier-space time-series phase-shifting; non-brain removal using BET; spatial smoothing using a Gaussian kernel of full-width-half-maximum 5mm.
Brain Images were normalized to standard space with the MNI 152 (average brain image at Montreal Neurological Institute) template using FLIRT (http://www.fmrib.ox.ac.uk/analysis/research/flirt) and resampled to 2 x 2 x 2 mm resolution. Data was band pass filtered ($0.01$Hz-$0.1$Hz) using a zero lag filter to avoid scanner drift and high frequency artifacts.
 
 %%%%%%%%%%%%%%%%%%%%%%%%
 \begin{figure}[h]
\centerline{
\includegraphics[width=.475\textwidth,clip=true]{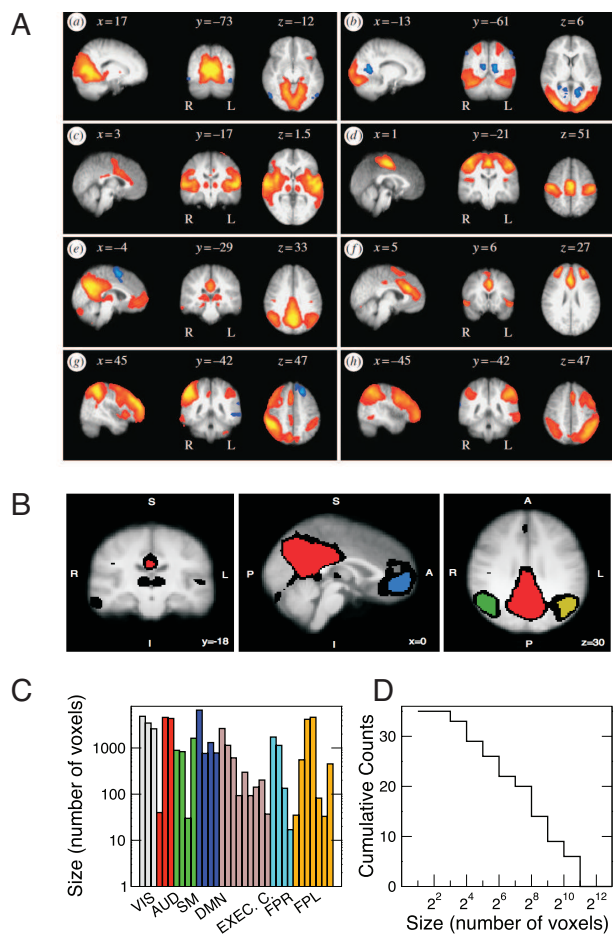}}
 \caption{Panel A: Spatial maps of the eight most relevant brain resting networks as described by Beckmann et al.\cite{beckmann2005}. Each map shows the locations of each RSN (shown in sagittal, coronal and axial views) where the coordinates refer to mm distances from the anterior commissure. Label VIS corresponds to visual; AUD  to auditory; SM to sensory motor; DMN to default model network; EXEC. C. to executive control; FPR and FPL to fronto-parietal right and left respectively.  
Panel B: Example (coronal, sagital, and axial views) of the four regions of interest extracted from the  DMN. The red region is composed of 6611 voxels, the blue region of 761, the green one of 1308, and the yellow region contains 780 voxels. Black voxels correspond to the ones in the original thresholded Z-map.   Bottom panels depict the sizes of the thirty-five clusters (C) studied here  as well as its cumulative size distribution (D).  }
\end{figure}

{\bf Choice of regions of interest.} It is known that the brain activity fluctuations at rest exhibit large-scale spatial correlations. The presence of these robust correlations is reflected on the coherent activity which determine the spatial domains of the RSN.  Therefore our analysis is focussed on the statistical analysis of the RSN fluctuations. At least since Beckmann et al.\cite{beckmann2005} Probabilistic Principal Component Analysis (PICA) is used to  identify the eight most relevant RSN. Each component corresponds to a characteristic time series, and its respective spatial Z-score map. Under a Gaussian/Gamma mixture model these Z-maps were thresholded in order to find the locations of the voxels which significantly contribute to each of the eight time-courses (see Fig. 6 in \cite{beckmann2005}) and used to define the clusters here. 
This is illustrated in  Fig. 1A, where the depicted regions correspond to the territory covered by each of the RSN extracted in \cite{beckmann2005} using ICA techniques. 
For each independent component Z-map we arbitrarily set a threshold that segment the map into isolated regions of different sizes (see Fig. 1B). 
The criteria to select regions is arbitrary, but the present results are independent of the selection criteria, as long as the regions belong to the same RSN. Alternatively, functional areas (such as Brodmann areas) can be used to define clusters of different sizes (as for a portion of the results in Fig. 3). We proceed by using a spatial mask for each of the eight networks to extract the time series of the BOLD fMRI time series. 
The masks, in Fig. 1, correspond to the eight most important RSN, namely the visual medial (box a) and lateral (b) cortical areas, the auditory (c), sensory motor (d), default mode (e), executive control (e), and the fronto-parietal right (g) and left (h) regions.   Each network  is comprised by a variable number of spatial clusters, each cluster  composed  by a variable number of voxels.
For instance the visual RSN (VIS) includes just three relatively large clusters, each one composed by thousand of voxels, contrasting with the Fronto-Parietal Left (FPL) network which involves  seven clusters with sizes ranging from a few up to thousands of voxels. Table 1 shows the thresholds used in each independent component and how many regions have been defined. The results presented in this paper are independent of the particular value of threshold used.
\begin{table}[h]
\begin{center}
\begin{tabular}{|c|c|c|c|c|c|c|c|c|}
\hline
RSN & Vis. 1& Vis. 2 & Aud. & Sens. M. & D.M. & E.C. & F.P.R & F.P.L  \\
\hline
 Threshold & 4 &3.3 & 2.4 &3.4 & 2.2& 2.7&3.2 &2.2 \\
\hline 
 \# regions &1 & 2&3 & 4&4 &9 &4 &8 \\ 
\hline 
\end{tabular}
\caption{ Z-threshold used in each independent component for defining the regions.}
\end{center}
\label{default}
\end{table}

\section{Results}
To analyze the noise properties,  we  look at the behavior of the variance and correlations under various manipulations of the size of the ensemble of voxels where these fluctuations occurs. This is a common strategy in other statistical physics problems  where very distinctive scaling behavior can be observed depending of the type of fluctuations the system is able to exhibit \cite{stanley}.

{\bf Anomalous scaling of the variance.}
We start by studying the fluctuations of the BOLD signal around its mean. The signal of interest,  for the  thirty-five RSN clusters, is defined as
\begin{equation}
{B_h}(\vec{x}_i,t)=
B(\vec{x}_i,t)-\frac{1}{N_H}\overset{N_H}{\underset{i=1}{\sum}} B(\vec{x}_i,t),
\end{equation}
where $\vec{x}_i$ represents the position of the voxel $i$ that belongs to the cluster $H$ of size $N_H$. 
These signals will be used to study the correlation properties of the activity in each cluster.

The mean activity of each $h$ cluster is defined as
\begin{equation}
\overline{B}(t)= \frac{1}{N_H}\overset{N_H}{\underset{i=1}{\sum}} B(\vec{x}_i,t),
\end{equation}

and its variance is defined as 
\begin{equation}
\sigma_{_{\overline{B}(t)}}^2=\frac{1}{T}\overset{T}{\underset{t=1}{\sum}}(\overline{B}(t)-\overline{\overline{B}})^2,
\end{equation}
where $\overline{\overline{B}}=\frac{1}{T}\overset{T}{\underset{t=1}{\sum}}\overline{B}(t)$ and $T$ the number of temporal points. Please notice that the average subtracted  in Eq. 1 is the mean  at time $t$  (computed over $N$ voxels) of the BOLD signals, not to be confused with the BOLD signal averaged over $T$ temporal points.

Since the BOLD signal fluctuate widely and the number $N$ of voxels in the clusters can be very large, one might expect that the aggregate of Eq. 1 obeys the law of the large numbers. If this was true, the variance of the mean field $ \sigma_{_{\overline{B}(t)}}^2$ in Eq. 3 would decrease with $N$ as $N^{-1}$. 
In other words one  would expect a smaller amplitude fluctuation for the average BOLD signal recorded in clusters (i.e., $\overline{B}(t)$)   comprised by large number of voxels compared with smaller clusters. However, the data in Fig. 2  shows otherwise, the variance of the average activity remains approximately constant over a change of four orders of magnitude in cluster' sizes. The strong departure from the $N^{-1}$ decay is enough to disregard further statistical testing. Nevertheless, we test a null hypothesis recomputing  the variance for artificially constructed clusters having similar number of voxels but composed of the randomly reordered $B_k(t)$ BOLD raw time series (panels in Fig. 2A denoted ``Shuffled''). As expected, in this case the variance  (plotted using squares symbols in Fig. 2B) obeys the $N^{-1}$ law (dashed line in Fig. 2B). 
 %%%%
\begin{figure}
\centerline{
\includegraphics[width=.5\textwidth,clip=true]{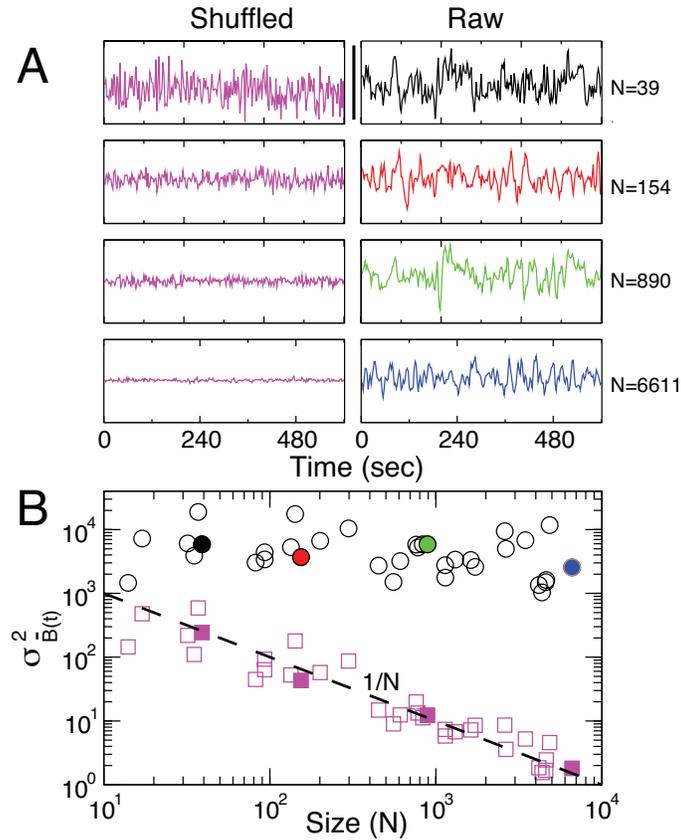}
}
\caption{Anomalous scaling of brain BOLD temporal fluctuations. Top panels show four examples of average BOLD time series (i.e., $\overline{B}(t)$  in Eq. 2) computed from clusters  of different sizes $N$. Note that while the amplitude of the raw BOLD signals (right panels) remains approximately constant, in the case of the shuffled data sets (left panels) the amplitude decreases drastically for increasing cluster sizes.
Panel B shows the calculations for the thirty five clusters (circles) plotted as a function of the cluster size demonstrating that variance is independent of the RSN's cluster size. The squares symbols show similar computations for a surrogate time series constructed by randomly reordering the original BOLD time series, which exhibit the expected $1/N$ scaling (dashed line). Filled symbols in Panel B are used to denote the values for the time series used as examples in Panel A.  }
 \end{figure}
 %%%%%
The variance of the average BOLD signal is directly proportional to the coordination between the voxels involved. In particular, under the hypothesis that the BOLD signal of voxel $k$, $B_{k}(t)$, is a stationary stochastic process (indexed by time $t$) with
$\mathbb{E}(B_{k}(t))=\mu_{k}$, and $\mathbb{V}ar(B_{k}(t))=\sigma_{k}^2$, the variance of the average signal is maximum in the case where there exist perfect coordination (i.e. all BOLD signals are perfectly synchronized). In this last case, the value of $\sigma_{_{\overline{B}(t)}}^2$ is equal to the mean value of the individual time variances defined as
 \begin{equation}
\tilde{\sigma}_{_{\overline{B}(t)}}^2:=\frac{1}{N}\overset{N}{\underset{k=1}{\sum}}\sigma_{k}^2.
\end{equation}
The inset of Fig. 3 (circles) shows that this maximum value does not depend on $N$, i.e. the mean value of the variance of the BOLD signal from a region does not depend on its size.
Now we ask how far from its maximum value is the observed variance of the BOLD average signal.
In particular, we compute the quotient
\begin{equation}
 q:=\frac{\sigma_{_{\overline{B}(t)}}^2}{\tilde{\sigma}_{_{\overline{B}(t)}}^2}
 \end{equation}
 for this purpose. As it is shown in Fig. 3 (empty circles) the value of $q$ decreases rather slowly with the size of the cluster.

In order to distinguish how much of the constancy of the variance demonstrated up until now is related with the fact that the time series belong to clusters that are independent components \cite{beckmann2005}  we repeated the scaling analysis using clusters defined by the Brodmann areas.
The results in Fig. 3 confirm the same anomalous scaling behavior demonstrated for the regions selected from the RSN networks, as shown by the values of $\tilde{\sigma}_{_{\overline{B}(t)}}^2$ and $q$ for the Brodmann areas (filled triangles). As before, we control the expected $1/N$ scaling for independent time series by computing the quotient $q$  for clusters of various sizes constructed  from  a random selection of voxels. This is shown by the filled square points in Fig. 3.
 %%%%%
 \begin{figure}[h]
\centerline{
\includegraphics[width=.5\textwidth,clip=true]{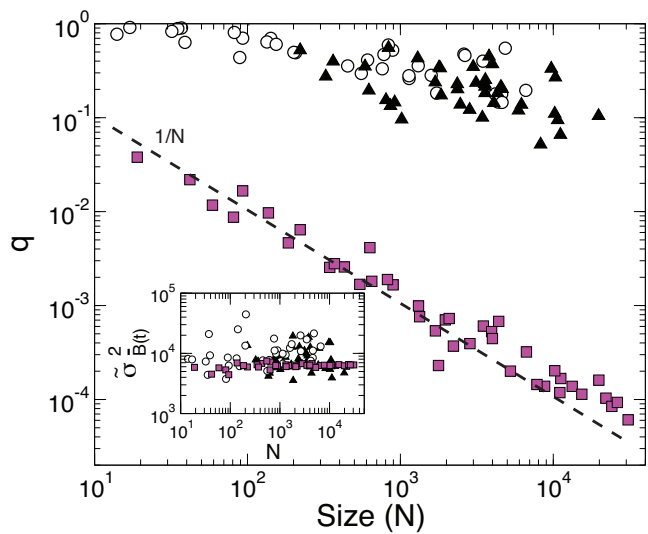}}
 \caption{The value of the quotient $q$,  expressing the measured variance relative to its maximum possible value, as a function of cluster size $N$.
Empty circles correspond to the 35 regions derived from the RSN (same as in Fig. 2) and filled triangles to the 41 Brodmann areas. The filled squares, obeying the $1/N$ scaling (dashed line), correspond to clusters of different sizes constructed from a random selection of voxels.The inset shows the average maximum of the BOLD signal variance of a cluster ($\tilde{\sigma}_{_{\overline{B}(t)}}^2$) as a function of $N$.  }
 \end{figure}
 
{\bf Temporal fluctuations and spatial correlations.} For spatio-temporal signals  the relationship between the temporal fluctuations of the average signal and its space correlation function is well defined \cite{randomprocesses}. In our case, for the normalized (see Appendix) BOLD signals, $Z_i(t)$ ($\mathbb{V}ar(Z_k(t))=1$ and $\mathbb{E}(Z_k(t))=0$), the  relationship is:
\begin{equation}
\sigma_{_{\overline{Z}(t)}}^2= \frac{1}{N}(1 + (N-1)\cdot \langle C \rangle).
\end{equation}
Where $\langle C \rangle$ is the mean spatial correlation,
\begin{equation}
\langle C \rangle=\frac{2}{N(N-1)}\overset{N-1}{\underset{i=1}{\sum}} \overset{N}{\underset{j=i+1}{\sum}} \rho_{i,j} ,
\end{equation}
$\rho_{i,j}$ the correlation between voxels $i$ and $j$, 
and $\sigma_{_{\overline{Z}(t)}}^2$ is the variance of the average signal (defined in Eq. 3). Eq. 6 shows that the variance of the mean activity depends on the size of the region, and on $\langle C \rangle$, which is determined by the shape of the correlation function, $C(r)$ (see Appendix for a formal discussion).

Eq. 6  suggest that it can be productive to investigate the correlations properties of the BOLD data. The point to clarify is whether the average spatial correlation $\langle C \rangle$ is constant throughout the entire recordings or, alternatively, its average value is the product of a combination of some instances of high spatial coordination intermixed with moments of dis-coordination. The relevance of this distinction, which will be further discussed latter,  is to establish up to which point correlations are dictated by the structural (i.e., fixed) connectivity or by the dynamics. 
In order to answer this question we study the mean correlation ($\langle C \rangle$) as a function of time for regions of interest of various sizes. In particular, we compute this value using Eq. 7 but estimating $\rho_{i,j}$ for non-overlapping periods of 10 temporal points.
 
Figure 4 shows the behavior of $\langle C \rangle$ over time for four different cluster's sizes. Notice that, in all cases, there instances of large correlation followed by moments of week coordination, as those indicated by the arrows in the uppermost panel.  We have verified that this behavior is not sensitive to the choice of the length of the window in which  $\langle C \rangle$  is computed (see the Appendix).
These bursts keep the variance of the correlations almost constant (i.e., in this example, there is a minor decrease in variance (by a factor of 0.4) for a huge increase in size (by a factor of 170). This peculiar behavior of the correlation is observed for any of the cluster sizes as shown in the bottom panel of Fig. 4 where the variance of $\langle C \rangle$  is approximately constant, despite the four order of magnitude increase in sizes.

The results of these calculations implies that independently of how large the size of the cluster considered, there is always an instance in which a large percentage of voxels are highly coherent and another instance in which each voxels activity is relatively independent.

A very metaphorical way to visualize the behavior of the correlations is to think of the patterns of spontaneous activity as ``clouds'' of relatively higher activity moving slowly throughout the brain's cortex.  Thus, the moments of large coordination shown in Fig. 4 correspond to the passage of a ``cloud'' throughout  the entire region, regardless of how large the region is. 
%%%%%%
\begin{figure}[h]
\centerline{
\includegraphics[width=.5\textwidth,clip=true]{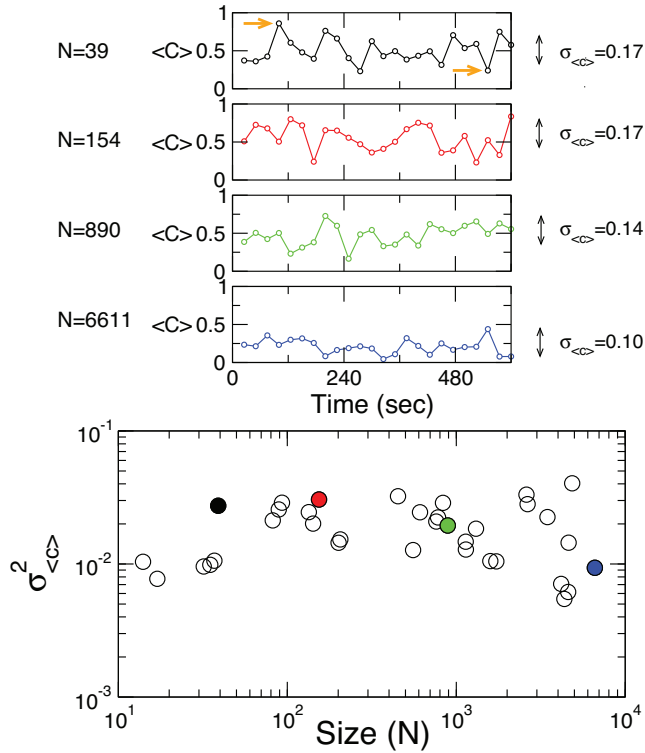}}
 \caption{Bursts of high correlations are observed at all cluster sizes, resulting in approximately the same variance, despite the four orders of magnitude change in the cluster size. The top panels illustrate representative examples of short-term mean correlation  $\langle C \rangle$ of the BOLD signals as a function of time for four sizes spanning four orders of magnitude.  The arrows show examples of two instances of highly correlated and weakly correlated activity, respectively. Bottom panel shows the variance of $\langle C \rangle$ as a function of cluster sizes. The four examples on the top traces are denoted  with filled circles in the bottom plot. }
 \end{figure}
 
{\bf Divergence of the correlation length.}
The results in the previous paragraphs indicate that the anomalous scaling of the variance can be related to dynamical changes in the correlations.   A straightforward  approach to understand the correlation behavior commonly used in large collective systems \cite{birds} is to determine the correlation length at various system's sizes. The correlation length is the average distance at which the correlations of the fluctuations around the mean crosses zero. It describes how far one has to move to observe any two points in a system behaving independently of each other. Notice that, by definition, the computation of the correlation length is done over the fluctuations around the mean, and not over the raw BOLD signals, otherwise global correlations may produce a single spurious correlation length value commensurate with the brain size. 

Thus, we start by computing for each voxel BOLD time series their fluctuations around the mean of the cluster that they belong. Recall the expression in Eq. 1:

\begin{equation}
{B_h}(\vec{x}_i,t)=
B(\vec{x}_i,t)-\frac{1}{N_H}\overset{N_H}{\underset{i=1}{\sum}} B(\vec{x}_i,t),
\end{equation}

where $B$ is the BOLD time series at a given voxel and $\vec{x}_i$ represents the position of the voxel $i$ that belongs to the cluster $H$ of size $N_H$. By definition
the mean of the BOLD fluctuations of each cluster vanishes,
\begin{equation}
\overset{N_k}{\underset{i=1}{\sum}} \overline{B_h}(\vec{x}_i,t)=0 \quad \quad \forall t.
 \end{equation}
Next we compute the average correlation function of the BOLD fluctuations
 between all pairs of voxels in the cluster considered, which are separated by a distance $r$:
\begin{equation}
\langle C_H(r)\rangle=<\frac{(B_H(\overrightarrow{x},t)-<B_h(\overrightarrow{x},t))>_t)(B_H(\overrightarrow{x}+r\overrightarrow{u},t)-<B_h(\overrightarrow{x}+r\overrightarrow{u},t)>_t)}{(<B_H(\overrightarrow{x},t)^2>_t-<B_H(\overrightarrow{x},t)>_t^2)^{1/2}(<B_H(\overrightarrow{x}+r\overrightarrow{u},t)^2>_t-<B_H(\overrightarrow{x}+r\overrightarrow{u},t)>_t^2)^{1/2}}>_{t,\overrightarrow{x},\overrightarrow{u}}
\end{equation}
where $\vec{u}$ is a unitary vector, and $\langle . \rangle_{w}$ represent averages over $w$.
 %%%
\begin{figure}[h]
\centerline{
\includegraphics[width=.3\textwidth,clip=true]{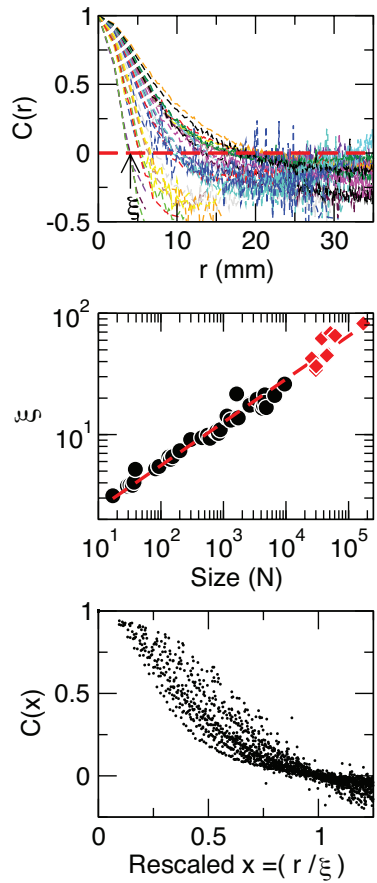}}
 \caption{Contrary to naive expectations, large clusters are as correlated as relatively smaller ones: the correlation length increases with cluster size.  Each line in the top panel shows the mean cross-correlation $C(r)$ of BOLD activity fluctuations as a function of distance $r$ averaged over all time series of each of the thirty five clusters shown in Fig. 1.  The correlation length $\xi$, denoted by the zero crossing of $C(r)$ is not a constant.
The middle panel  shows, in double log plot, the functional dependence $\xi \sim d N ^{1/3}$, i.e., $\xi$ grows linearly with the average cluster' diameter $d$ for all the thirty five clusters (filled circles).  The rightmost points (diamonds) corresponding to the  $\xi$ values computed for each of the eight RSN without any partitioning shows that the correlation length keep increasing up to the size of the brain (the dotted line is a guide to the eye with slope 1/3)
The scale invariance is graphically illustrated by the bottom panel, where all $C(r)$ data are replotted after rescaling the horizontal axis as $x=r/\xi$, showing a good overlap. Note that a perfect collapse of these curves can not be expected because of the severe anisotropy, imposed by the brain anatomy, affecting the estimation of the distance $r$.}
 \end{figure}
 %%%%%%%%%%%%%%%%%%%%
The typical form we observe for $C(r)$  is shown in the top panel of Fig. 5. The first striking feature to note is the absence of a unique $C(r)$ for all clusters. Nevertheless, they are qualitatively similar, being at short distances close to unity, to decay as $r$ increases, and then becoming negative for longer voxel-to-voxel distances. Such behavior indicates that within each and any cluster, on the average, the fluctuations around the mean are strongly positive at short distance and strongly anti-correlated at larger distances, whereas there is no range of distance for which the correlation vanishes.

The most notorious result is the fact that correlations decay with distance slower in larger clusters than in relatively smaller clusters, giving rise to the family of curves shown in Fig. 5 (top panel). This is condensed in the calculation of the correlation length $\xi$,  which is the zero of the correlation function, $C(r =\xi) = 0$ (as in the example shown by the arrow in Fig. 5, top). The correlation length diverges with the size of the cluster, as demonstrated in the middle panel of Fig. 5. This divergence extends up to the size of the brain, as shown by the $\xi$ values (red squares in middle panel of Fig. 5) computed for the eight unpartitioned RSN. Note that while the existence of a zero crossing in $C$ is warranted by the subtraction of the mean cluster activity (in Eq. 8), its divergence with cluster size is not.
%%%%
\begin{figure}[h]
\centerline{
\includegraphics[width=.3\textwidth,clip=true]{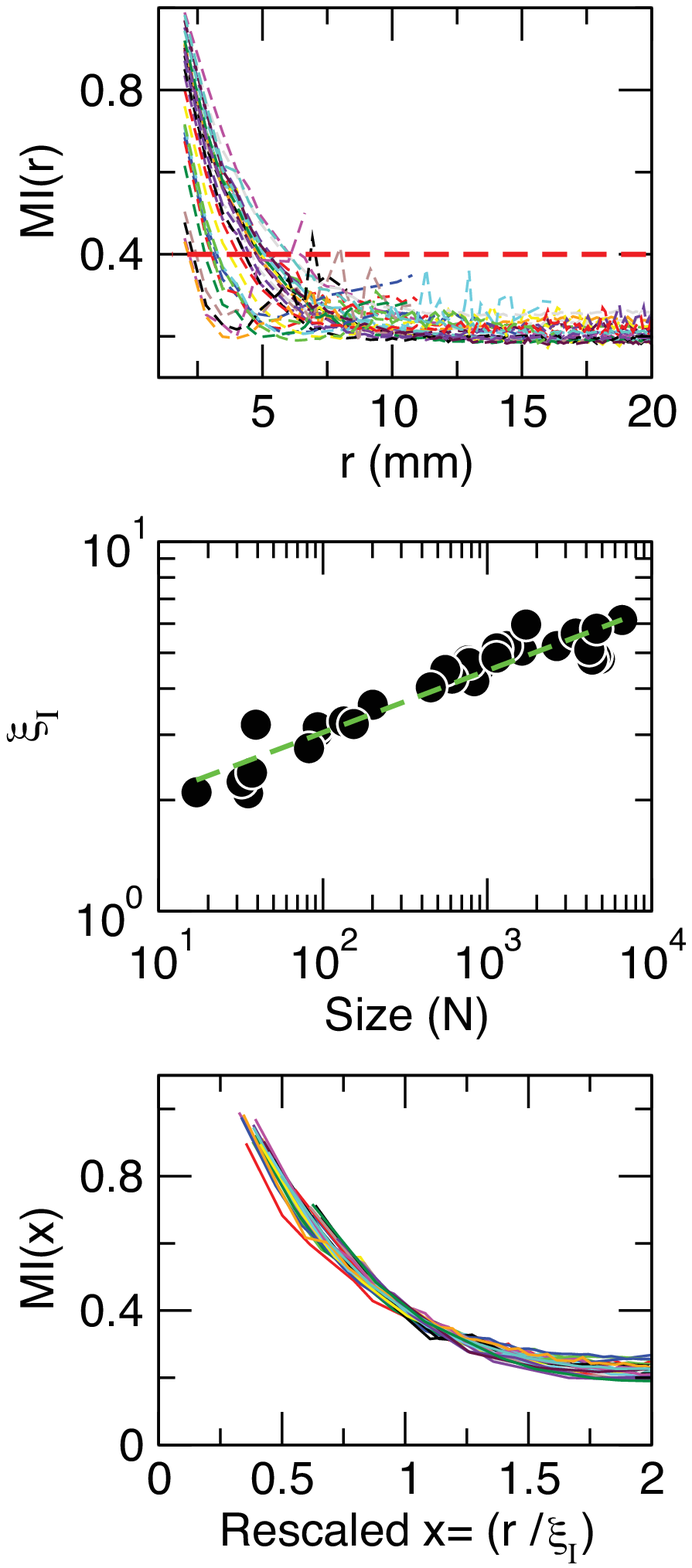}}
 \caption{ Mutual information increases with cluster size.  Each line in the top panel shows the mutual information $MI(r)$ as a function of distance $r$ averaged over all time series of each of the thirty five clusters shown in Fig. 1.  The length at which  $MI(r)$ decreased to a given value (red line in the top panel) denoted as $\xi_I$,  is an increasing function of the size of the cluster (middle panel).  The bottom panel  illustrates the good data collapse after rescaling the horizontal axis as $x=r/\xi_I$.  }
 \end{figure}
 %%%%%%

{\bf Mutual Information.}
Although the present observations can be appropriately described solely in terms of correlations, the same concept can be also casted in terms of information measures, which are often used to estimate the degree of coherence between regions or neural structures.  The mutual information between any two $X$ and $Y$ time series from different brain voxels is defined as:
\begin{equation}
MI(X;Y) = H(X) - H(X | Y)
\end{equation}
where $H(X)$ is the entropy of $X$ and $H(X|Y)$ is the entropy of $X$ given $Y$ computed as usual \cite{info}. In principle, given the behavior observed for the correlations, these information measures should exhibit scale-invariant scaling as well. This is confirmed by the results in Fig. 6, which demonstrate that the average mutual information is not affected by the size of the cluster considered, since information decays slower in larger clusters. This analysis shows that, as was shown for the correlation, the information length (determined here for an arbitrary threshold value of 0.4 bits) diverges with the size of the 
of the clusters.

\section{Discussion}

In this work, key statistical properties of the brain BOLD signal variability were investigated. The results are relevant to the understanding of the brain spontaneous activity fluctuations in health and disease. The three most relevant findings that we may discuss are: 
\begin{itemize}

\item{the variance of the average BOLD fluctuations computed from ensembles of widely different sizes remains constant, (i.e., anomalous scaling);}

\item{the analysis of short-term correlations reveals bursts of  high coherence between arbitrarily far apart voxels indicating that the variance' anomalous scaling has a dynamical (and not structural) origin; }

\item{the correlation length measured at different regions increases with region's size, as well as its mutual information.}
 
\end{itemize}

Concerning the constant variance of the BOLD activity, the present results imply that the usual framework in which  the BOLD signal and noise are discussed need to be reconsidered. For instance, it is commonplace to consider that the non-coherent part of the activity (i.e., the noise) can be averaged out by enlarging the spatial (i.e., more voxels) or temporal (i.e., more samples) scale. The presence of anomalous scaling implies that signal and noise in the brain are at least ill defined and that filtering by averaging (to improve its quality) signals with anomalous variance, by definition, can be anomalous as well.
The anomalous scaling also has implication for the monitoring of the RSNs activity, a topic that has received wide attention recently for its potential to track healthy or pathological conditions. The results here imply that, under these anomalous conditions, the signal of a few voxels can be, asymptotically, as representative and informative as the average of the entire RSN. It need to be noted, that the anomalous scaling discussed here due to the emergence of collective dynamics is not new, Kaneko \cite{Kaneko} demonstrated the breach of law of large numbers in numerical models more than two decades ago.
 
The second finding, showing that the observed dynamical short-term changes in the correlations drives up the variance, is relevant for the interpretation of the brain functional connectivity.  The evaluation of functional connectivity between regions often uses the average correlation, and the results in Fig. 4 show that, despite the relatively weak average functional connectivity values, it can be instances in which the correlation reaches high levels. In other words, under the demonstrated  anomalous scaling conditions, the usual pairwise measures has inherent limitations for the proper interpretation of these collective states. In passing, it need to be noted that these instances of high coherence were recently confirmed using a different method, which demonstrate avalanches of activity encompassing relatively large regions of each RSN \cite{Enzo2012}. Of course, the role of these epochs of transient synchronous states in driving perception, awareness and consciousness are consistent with the views championed by Varela and coworkers more than a decade ago \cite{varela} as discussed recently \cite{werner}.

The third result concerning the divergence of the correlation length with increasing cluster size is perhaps the most telling one, because is in contrast with the prevailing viewpoints about brain functional connectivity. Indeed it is implicit in the interpretation of functional connectivity studies the notion that brain activity {\it propagates} between and across brain regions. However, for such propagating waves a constant correlation length (i.e. its wavelength) is always expected, which is not what it is consistently found in the present data.  The divergence of correlations with size  (and its associated anomalous scaling) suggests, in addition, that our current mathematical approaches to  model cortical dynamics could be ill-fated. The issue is that most of the large scale models (for superb reviews see \cite{sporns,deco}) are defined by an adjacency matrix specifying the ``structural connectivity'' between a large number of regions and some kind of neural dynamics assigned to each node (i.e., cortical region). Lets imagine that such model is scaled up by increasing the number of regions an order of magnitude, while the correlation length of the activity fluctuations is measured as in the experiments here.  A reasonable  conjecture is that  current large-scale brain models would have problems to replicate the present findings, since anomalous scaling only appears at criticality (discussed below) while current models are purposely tuned to the ordered regime.

Finally, an important question is concerned with the origin of the statistical properties unveiled in this work.  We suggest that a candidate explanation which is able to unify all the observations presented here can be found in the context of critical phenomena \cite{stanley,bak,kim}. It is well known that dynamical systems composed of very large number of  interacting nonlinear elements, under some conditions, exhibit emergent collective behavior with ubiquitous properties \cite{more}. Examples of emergent phenomena sharing common features are the collective dynamics of birds in a flock \cite{birds}, spins of a magnet\cite{stanley}, water molecules in the atmosphere \cite{peters}, peopleÕs financial decisions \cite{econ} or ants traffic in a foraging swarm \cite{ants,millonas}.  In all these cases, each agent in isolation may have its own stereotypical behavior,  but when placed to interact in very large numbers, and under certain conditions, the entire system will drift toward a type of collective dynamics which lies in between complete order and complete disorder. At this point (known as an order-disorder phase transition \cite{stanley,dante}) the collective dynamics of the system exhibit distinctive universal properties. Amongst them, the most significant common features include the divergence of correlations, the anomalous scaling, and the presence of moments of high coordination seen here for the RSN fluctuations. Since the emergence of these properties require conditions near an order-disorder phase transition, its observation it is often considered a {\it distinctive signature} of critical dynamics, as reported recently by Cavagna et al. for sterling flock dynamics \cite{birds}.  In particular, it is known that only near a critical point $\xi$ can grow with system size, where the collective global effects overcomes the individuals dynamics, resulting in the emergence of correlated domains of arbitrary size, where information propagates equally well up to the size of the entire system.

In summary, the analysis of the BOLD' fluctuations of the resting brain shows anomalous statistical properties, bursts of highly correlated states and divergence of correlation length, which are dynamical properties known to be found only near a critical point of a phase transition. These findings are fully consistent with previous reports of large-scale brain critical dynamics \cite{Kitz,Dante-2009,Expert,dante,Enzo2012} and may be one answer to the question in the title in the sense that brain noise corresponds, rigorously speaking,  to the type of (spatial and temporal) fluctuations only observed in systems near criticality. This view may be relevant for the interpretation of the role of fluctuations and variability in brain function in health and disease.

\section{Appendix}
Additional information is provided here to supplement the main results. The first item is concerned with the robustness of the short-term correlations presented in Fig. 4. The second point deals with the generality of the divergence of correlations and the last  one discuss formally the presence of long-range correlations in the fMRI data.
 
\emph{Short-term correlations.} As discussed in Fig. 4, the presence of bursts of high and low correlations observed throughout  clusters of very different size is the dynamical base for the violation of the law of the large numbers. It is then important to demonstrate that the estimation of the  short-term correlations' variance is robust. For that, we recomputed the results in Fig. 4 for various window  lengths. This is presented in Fig. 7 which shows that the variance of $\langle C \rangle$ is independent of $N$ regardless of the window length at which it is estimated.  
%%%%%
\begin{figure}[h]
\centering{
\includegraphics[width=.5\textwidth]{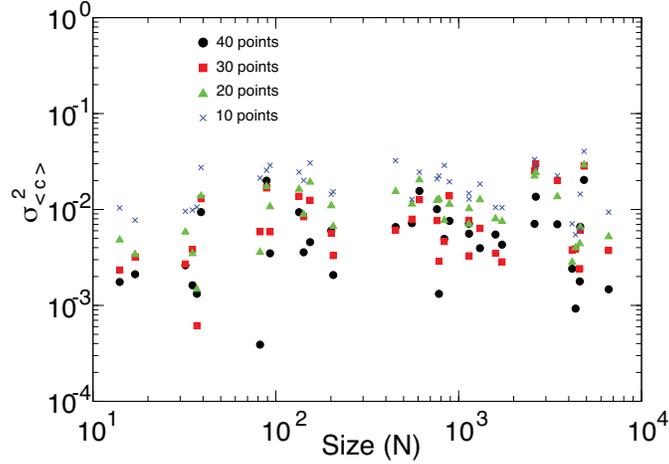} }
\caption{The variance of the mean short-term spatial correlation $\langle C \rangle$ (already shown in Fig. 4)  is independent of the cluster' sizes $N$ regardless of the window length (10 to 40 time steps) at which it is estimated.} 
\end{figure}
%%%%%

\emph{ $\xi$ scaling.} The divergence of correlation length discussed in Fig. 5 predicts a  functional dependence $\xi \sim d N ^{1/3}$, i.e., $\xi$ grows linearly with the average cluster' diameter $d$. The results in Fig. 8, obtained from the analysis of fMRI data from four different subjects, confirm such scaling relation.

\begin{figure}[h]
\centering{
\includegraphics[width=.5\textwidth]{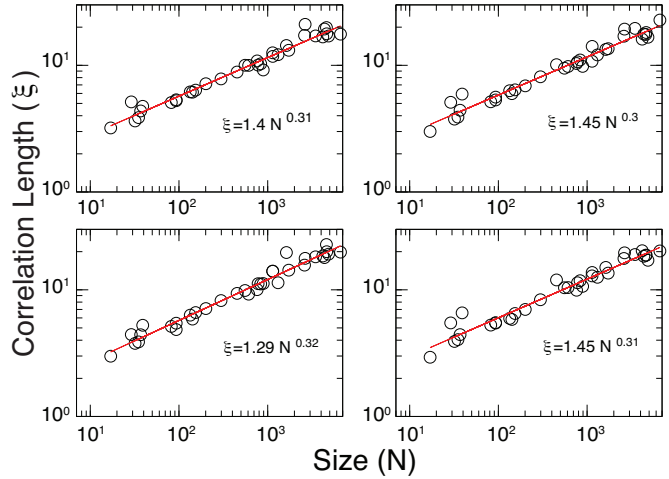} }
\caption{Estimation of the  correlation length $\xi$ divergence. In all cases the results are very close to the scaling found in Fig. 5;  $\xi \sim d N ^{1/3}$.  }
\end{figure}

\emph{Long-range correlations.} In spatio-temporal data it is well known the relationship between the temporal fluctuations of a mean magnitude and the space correlation function. Let suppose we want to study a brain region (our clusters in the main text) of $N$ voxels. Denote a voxel  of the region as $i$ which is characterized by its position in space  ($\overrightarrow{r}_i$), and by its dynamics represented in the BOLD signal ($B_i(t)$).
In addition, to simplify the notation we are going to work  here with normalized BOLD signals,
\begin{equation}
Z_i(t)\equiv \frac{B_i(t)- \overline{B_i}}{ \sigma_i}, 
\end{equation}
where $ \overline{B_i}=\frac{1}{T}\underset{h=1,\dots,T}\sum B_i(t)$ and $\sigma_i^2=\frac{1}{T}\underset{h=1,\dots,T}\sum (B_i(t)- \overline{B_i})^2$. Each voxel signal ($Z_i(t)$) has now zero mean and variance one. The average signal over the region, which is:  
\begin{equation}
\overline{Z}(t)=\frac{1}{N}\overset{N}{\underset{i=1}{\sum}}Z_i(t),
\end{equation}

fluctuates in time.  Our interest here are the fluctuations of $\overline{Z}(t)$.  It can be shown, using the definition of the variance of a sum of random variables, that 
the variance of the average signal of the cluster is:
\begin{equation}
Var(\overline{Z})= \frac{1}{N}(1 + (N-1)\langle C \rangle), 
\end{equation}
where $\langle C \rangle$ is the mean spatial correlation,
 $Var(\overline{Z})=\frac{1}{T} \overset{T}{\underset{t=1}{\sum}} (\overline{Z}(t)-\overline{\overline{Z}})^2$ 
 and  $\overline{\overline{Z}}=\frac{1}{T} \overset{T}{\underset{t=1}{\sum}} \overline{Z}(t)$.

Since we are interested also on how correlations affect variance, let consider some cases.  If there exist null variability between all the voxels in the region, that is all voxels of the region do exactly the same in time,  the left term of Eq. 14 remains equal to one no matter the size ($N$) of the region is.  In any other case $Var(\overline{Z})$ will be less than one.  
The variance of the mean activity depends on the size of the region, and on $\langle C \rangle$,   which is determined by the shape of the correlation function, $C(r)$. 
Therefore, in order to understand the asymptotic behavior of $Var(\overline{Z})$ with $N$ we need to make some hypothesis over $C(r)$.

First,  the mean correlation, 
\begin{equation}
\langle C \rangle=\frac{2}{N(N-1)} \underset{i<j}\sum cor(Z_i,Z_j),
\end{equation}
is approximated by its continuous version
\begin{equation}
\langle C \rangle \approx \frac{2\pi}{V} \int_{0.5}^{r^*} C(r) r^2dr,
\end{equation}
where $r^{*}$ is the radius of the spherical region under study, and $V$ its volume. Now, we discuss some hypothesis about the \textit{asymptotic} behavior of $C(r)$. For example, if
 there exist an exponential decay,
 \begin{equation}
C(r) \sim  e^{-\lambda r},
 \end{equation} 
 then the mean correlation satisfies:
  \begin{equation} 
\langle C \rangle \sim N^{-1}.
 \end{equation} 
 In the case where long-range correlations are present, 
  \begin{equation}
C(r) \sim   \frac{1}{r^{\alpha}},
 \end{equation} 
 the mean correlation satisfies:
  \begin{equation} 
\langle C \rangle \sim N^{-\alpha/3}.
 \end{equation} 
 Putting all together in Eq. 21, the spatial decay of the fMRI data correlations will be given by the product of $N$ by $Var(\overline{Z})$ as a function of $N$, leading to two very different asymptotic statistical behavior:
  \begin{equation}
 \begin{split}
 \begin{cases} 
 & \mbox{For long range correlations} \quad \quad NVar(\overline{Z}) \sim N^{1-\alpha/3}\\
 & \mbox{For short range correlations} \quad \quad  NVar(\overline{Z}) \sim k. \\
\end{cases}
\end{split}
\end{equation}
Fig. 9 shows $N.Var(\overline{Z})$ as a function of $N$ for brain data. The straight line in the log-log plot confirm that in the brain data there exist long range correlations. In particular, we obtain a exponent $\alpha=0.9$ (for $C(r) \sim   \frac{1}{r^{\alpha}}$) which agrees very well with the result recently obtained by Expert et al. \cite{Expert}. For completeness  we plot also the results of numerical calculations using an exponential correlation function which clearly depart from the brain data.  

\begin{figure}[h]
\centering{
\includegraphics[width=.5\textwidth]{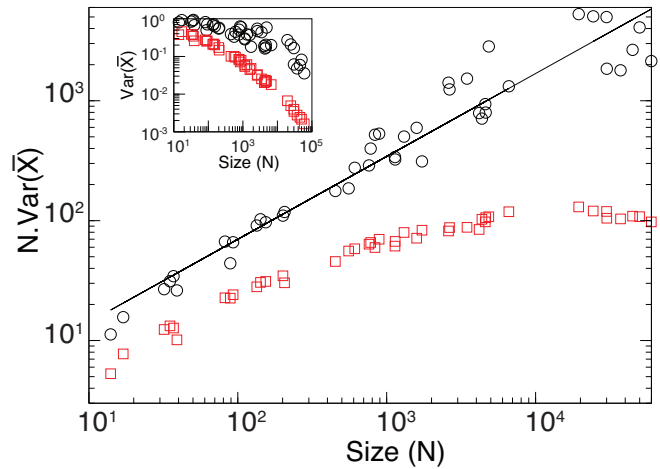}  }
\caption{Long-range correlations. Black circles correspond to brain data and red squares to the results from an exponential interaction model with the same geometry for each cluster. In the exponential case as $N$ grows the average correlation converges, meanwhile for the brain data it continues growing demonstrating the presence of long-range correlations in the data. The black line corresponds to a power law fit $y=k x^{-0.7}$. From Eq. 21 we obtain an exponent $\alpha=0.9$. The inset corresponds to the variance of the mean activity as a function of $N$. }\label{fig1}
\end{figure}

\section*{Acknowledgments}
Work supported by NIH NINDS (USA), grant NS58661, by Consejo Nacional de Investigaciones Cient\'ificas y Tecnol\'ogicas (CONICET) (Argentina), by the  Spanish Ministerio de Economia y Productividad (previously Ministerio de Ciencia y Tecnologia) (Spain) and by European Funds - FEDER, grant SEJ2007-62312. We thank Prof. Pedro Montoya and M. Mu\~noz (UIB, Mallorca, Spain) for discussions and help in data acquisition, and E. Tagliazucchi,  P. Balenzuela , A. Haimovici  (UBA, Argentina) and  L. Hess, A. Tardivo, A. Yodice  (UNR, Argentina) for continuous discussions.

\end{document}